\documentclass[prd,10pt, tightenlines, twocolumn, nofootinbib]{revtex4-1}
\input epsf.tex
\epsfclipon

\usepackage[utf8]{inputenc}
\usepackage{tikz-cd}
\usepackage{cancel}
\usepackage[normalem]{ulem}
\usepackage{amsmath,amsthm}
\usepackage{amsfonts}
\usepackage{mathtools}
\usepackage{wasysym}
\usepackage[mathscr]{eucal}
\usepackage[sans]{dsfont}
\usepackage{tensor}
\usepackage{bigints}
\usepackage{calc}
\usepackage{makecell}
\usepackage{bbm}
\usepackage{amsmath}
\usepackage{bm}
\usepackage{verbatim}
\usepackage{hyperref}
\usepackage{setspace}

\newcommand{\be}{\begin{equation}}
\newcommand{\ee}{\end{equation}}
\newcommand{\bea}{\begin{eqnarray}}
\newcommand{\eea}{\end{eqnarray}}

\usepackage{amsmath,amssymb,lmodern}
\usepackage{xcolor}
\begin{document}

\singlespacing

\title{Dynamical Henneaux-Teitelboim Gravity}

\author{Emma Albertini}
\email{emma.albertini17@imperial.ac.uk}
\affiliation{Imperial College London}

\author{Kyle Barnes}
\email{kyle\_barnes@alumni.brown.edu}
\affiliation{Brown University}

\author{Gabriel Herczeg}
\email{gabriel\_herczeg@brown.edu}
\affiliation{Brown University}

\date{\today}

\begin{abstract} 
We consider a modified gravity model which we call ``dynamical Henneaux-Teitelboim gravity" because of its close relationship with the Henneaux-Teitelboim formulation of unimodular gravity.  The latter is a fully diffeomorphism-invariant formulation of unimodular gravity, where full diffeomorphism invariance is achieved by introducing two additional non-dynamical fields: a scalar, which plays the role of a cosmological constant, and a three-form whose exterior derivative is the spacetime volume element. Dynamical Henneaux-Teitelboim gravity is a generalization of this model that includes kinetic terms for both the scalar and the three-form with arbitrary couplings. We study the field equations for the cases of spherically symmetric and homogeneous, isotropic configurations. In the spherically symmetric case, we solve the field equations analytically for small values of the coupling to obtain an approximate black hole solution. In the homogeneous and isotropic case, we perturb around de Sitter space to find an approximate cosmological background for our model.
\end{abstract}

\maketitle


\section{Introduction}
\subsection{Overview}
Cosmic inflation solves a number of fundamental problems in cosmology by positing an early period of rapid expansion driven by a constant part of the stress energy tensor---it provides a framework for explaining why the observable universe is nearly flat and isotropic \cite{guth1981inflationary, kleban2016inhomogeneous, east2016beginning}, and gives a mechanism for large-scale structure to form from density perturbations in the early universe \cite{liddle2000cosmological}. Meanwhile, the present cosmological epoch is dominated by dark energy that is usually attributed to a cosmological constant, albeit a vastly smaller one \cite{caldwell1998cosmological, ratra1988cosmological}. In light of these facts, as well as the recent controversy over the Hubble tension \cite{di2021realm, knox2020hubble}, and the long-standing cosmological constant problems \cite{weinberg1989cosmological, adler1995vacuum, padilla2015lectures}, simple modifications of general relativity where the cosmological constant is promoted to a dynamical field are worthy of consideration. 

One route to accomplishing this is to consider modified gravity models where the Bianchi identities no longer imply conservation of the stress energy tensor, but only the weaker condition that the divergence of the stress tensor is a total derivative. A minimal modification of general relativity in which this can be achieved is unimodular gravity \cite{josset2017dark}. In the standard formulation of unimodular gravity, the spacetime volume element is constrained to be equal to a prescribed background density, with the result that the equations of motion become the \emph{trace-free} Einstein equations
\be
R_{\mu\nu} - \frac{1}{4}R g_{\mu\nu} = \kappa(T_{\mu\nu} - \frac{1}{4}Tg_{\mu\nu}), \qquad \kappa = 8\pi G_N.\label{TFE}
\ee
Taking the divergence of \eqref{TFE} and using the contracted Bianchi identity $\nabla^\mu R_{\mu\nu} = \frac{1}{2}\nabla_\nu R$, leads to
\be 
\nabla^\mu T_{\mu\nu} = \frac{1}{4}\nabla_\nu(R + \kappa T).
\ee 
If one \emph{imposes} conservation of the stress tensor, then we learn that $\Lambda \equiv \frac{1}{4}(R + \kappa T)$ is a constant, which is just the pure-trace part of the Einstein equations, in which case unimodular gravity is classically equivalent to general relativity. However, if we are willing to consider non-conserved stress tensors, then unimodular gravity is physically distinct from general relativity even at the classical level.\footnote{For a recent review of the sometimes subtle differences between unimodular gravity and general relativity, see \cite{carballo2022unimodular}.} Some recent proposals for modified gravity models that draw on ideas from unimodular gravity and topological field theory lead to the same type of non-conservation of the stress tensor, and consequent fluctuations in the effective cosmological constant\cite{Alexander:2018tyf,alexander2020chiral, alexander2022black}. Here, we promote the cosmological constant to a dynamical field by considering a minimal modification of a fully diffeomorphism invariant, background independent reformulation of unimodular gravity due to Henneaux and Teitelboim \cite{henneaux1989cosmological} and we study the implications for spherically symmetric black holes and homegeneous, isotropic cosmology.  

The paper is structured as follows: in section \ref{UM}, we give a brief review of unimodular gravity; in section \ref{DHT}, we introduce dynamical Henneaux-Teitelboim gravity; section \ref{BH} is devoted to the study of spherically symmetric black holes, and in section \ref{FRW} we take a first look at the implications of dynamical Henneaux-Teitelboim gravity for a simple homogeneous, isotropic cosmological model. 
\subsection{Unimodular Gravity}\label{UM}
The action for unimodular gravity is simply the Einstein-Hilbert action where the spacetime volume element $\sqrt{-g}$ has been replaced with a fixed background density $\epsilon$:
\be 
S_{\mbox{\tiny UMG}} = \int_{\mathcal{M}}\epsilon\left(\frac{1}{2\kappa}R(g) + \mathcal{L}_M(g,\psi)\right)d^4x, \label{UMaction}
\ee 
where $g_{\mu\nu}$ is a metric satisfying $\sqrt{-g} = \epsilon$, and $\psi$ represents any matter fields that are present. Varying \eqref{UMaction} with respect to $g^{\mu\nu}$ leads to the trace-free Einstein equation \eqref{TFE}. The introduction of the background density $\epsilon$ breaks diffeomorphism invariance down to \emph{transverse} diffeomorphism invariance---i.e., invariance with respect to diffemorphisms generated by vector fields $\xi$ satisfying \be
\mathscr{L}_{\xi}(\epsilon \,d^4 x) = 0.
\ee 

In order to preserve full diffeomorphism invariance while retaining the spirit of unimodular gravity, Henneaux and Teitelboim \cite{henneaux1989cosmological} considered the action\footnote{Smolin has shown in \cite{smolin2009quantization} that certain aspects of the cosmological constant problem can be resolved using path integral methods based on the action \eqref{HTaction}.} 
\begin{equation}
    S_{\mbox{\tiny HT}} = \frac{1}{\kappa}\int \mathbf{\mathbf{V\!ol}}(g)\, \left(\frac{1}{2}R(g) - \varphi + \kappa \mathcal{L}_M(g,\psi)\right) + \varphi \, d\mathbf{H}, \label{HTaction}
\end{equation}
where $\varphi$ is a scalar, $\mathbf{H}$ is a three-form, and $\mathbf{V\!ol}(g) = \sqrt{-g}\,d^4x$ is the volume form associated with $g_{\mu\nu}$.   Varying with respect to $g^{\mu\nu}$ gives the full set of Einstein's equations with $\varphi$ playing the role of the cosmological constant:
\be 
R_{\mu\nu} - \frac{1}{2}R g_{\mu\nu} + \varphi g_{\mu\nu} = \kappa T_{\mu\nu}. \label{EFE1}
\ee
On the other hand, varying with respect to $\mathbf{H}$ leads to $d\varphi = 0$, so the gravitational equations of motion are equivalent to the usual Einstein equations. Finally, varying with respect to $\varphi$ gives $\mathbf{V\!ol}(g) = d\mathbf{H}$, so while $\mathbf{H}$ is varied in the action, it can be thought of as a potential for the spacetime volume in much the same way that $\epsilon$ fixes the spacetime volume in the standard formulation of unimodular gravity. Since there are no background structures, the action \eqref{HTaction} is invariant under the full group of diffeomorphisms. There is also a gauge symmetry under $\mathbf{H} \to \mathbf{H} + d\boldsymbol{\Omega}$ for any two-form $\boldsymbol{\Omega}$. 

In addition to restoring background independence and full diffeomorphism invariance, the Henneaux-Teitelboim formulation of unimodular gravity automatically has a conserved stress tensor, unlike the standard formulation. However, if one introduces additional terms into the Lagrangian that couple the new fields $\varphi$ and $\mathbf{H}$ to the metric, this may no longer be the case. As we shall see in the following section, introducing kinetic terms for both $\varphi$ and $\mathbf{H}$ is sufficient to violate conservation of the stress tensor just enough to induce fluctuations in the effective cosmological constant $\varphi$, although including a kinetic term for just one of them is not.

\section{Dynamical Henneaux-Teitelboim gravity}\label{DHT}
We now consider ``dynamical Henneaux-Teitelboim gravity," defined by the action
\begin{equation}
    S_{\mbox{\tiny DHT}} = S_{\mbox{\tiny HT}} - \alpha \! \int d\mathbf{H} \wedge *d\mathbf{H} -\beta \! \int d \varphi \wedge *d \varphi,\label{dynamical action}
\end{equation}
which enjoys the same symmetries as the Henneaux-Teitelboim formulation of unimodular gravity---it is background independent and fully diffeomorphism-invariant, and it retains the gauge symmetry $\mathbf{H} \to \mathbf{H} + d\boldsymbol{\Omega}$ for any two-form $\boldsymbol{\Omega}$. Here, and for the remainder of the article we use geometrized units $c = G_N = 1$, which renders $\kappa$ dimensionless, and gives the action dimensions of $\textrm{(length)}^2$. As a result, $\alpha$ acquires dimensions of $\textrm{(length)}^{-2}$, while $\beta \sim \textrm{(length)}^4$.

Varying \eqref{dynamical action} with respect to $g^{\mu\nu}$ leads to the Einstein field equations with the \emph{dynamical} field $\varphi$ playing the role of the cosmological constant, and where the stress tensor gets contributions from the old matter Lagrangian $\mathcal{L}_{M}(g,\psi)$, as well as the kinetic terms for $\mathbf{H}$ and $\varphi$: 
\begin{gather}
R_{\mu\nu} - \frac{1}{2}R g_{\mu\nu} + \varphi g_{\mu\nu} = \kappa T_{\mu\nu},\label{EFE2} \\
T_{\mu\nu} = T^{(\psi)}_{\mu\nu} + T^{(\varphi)}_{\mu\nu} + T^{(H)}_{\mu\nu}, \label{total stress} 
\end{gather}
where 
\bea  
T^{(\psi)}_{\mu\nu} &=& -\frac{2}{\sqrt{-g}}\frac{\partial \mathcal{L}_M}{\partial g^{\mu\nu}} + g_{\mu\nu}\mathcal{L}_M, \label{matter stress} \\
T^{(\varphi)}_{\mu\nu} &=& -\beta\left(\nabla_\mu \varphi \nabla_\nu\varphi- \frac{1}{2} g_{\mu\nu} \nabla^\rho \varphi \nabla_\rho\varphi\right), \label{phi stress} \\
T^{(\mathbf{H})}_{\mu\nu} &=& -\alpha\frac{\tilde{Q}^2}{|g|}g_{\mu\nu}, \label{H stress}
\eea   
with $\mathbf{Q} = d\mathbf{H}$, and $\tilde{Q} =\frac{1}{4!} Q_{abcd}\epsilon^{abcd}$.
The trace-free part of the gravitational equation \eqref{EFE2} is just the trace-free Einstein equation \eqref{TFE}, which, as we noted in the introduction, implies that the divergence of the stress tensor is a total derivative $\nabla^\mu T_{\mu\nu} = \frac{1}{4}\nabla_\nu(R + \kappa T)$. On the other hand, the pure trace part of \eqref{EFE2} is simply $\varphi = \frac{1}{4}(R + \kappa T)$. Putting these together gives 
\be  
\nabla_\nu \varphi = \nabla^\mu T_{\mu\nu}.
\ee
So we see that fluctuations in $\varphi$ are sourced by non-conservation of the stress tensor of precisely the type we have alluded to. As for the scalar and three-form equations of motion, we can vary \eqref{dynamical action} with respect to $\varphi$ to obtain 
\begin{equation}
    d\mathbf{H} = \mathbf{V\!ol}(g)- 2 \beta d*d \varphi, \label{delta phi}
\end{equation}
while varying \eqref{dynamical action} with respect to $\mathbf{H}$ gives
\be 
d\varphi -2 \alpha \, d*d\mathbf{H}=0. \label{delta H}
\ee 
Taking the Hodge star of \eqref{delta phi} and inserting the result into  \eqref{delta H} leads to
\be\label{ca5}
 d(\varphi +4\alpha \beta \square \varphi)  = 0,
\ee
while taking the Hodge star of \eqref{delta H} and plugging into \eqref{delta phi} gives
\be 
    d(\mathbf{H} +4 \alpha \beta d^\dagger d \mathbf{H})= \mathbf{V\!ol}(g),
    \\ \label{ca6}
\ee
where $d^\dagger \equiv *\,d\,*$. From equations \eqref{ca5} and \eqref{ca6}, we see that if either $\alpha = 0$ or $\beta = 0$, then we recover ${d\mathbf{H} = \mathbf{V\!ol}(g)}$ and $d\varphi = 0$, and the model is equivalent to the Henneaux-Teitelboim formulation of unimodular gravity. However, when $\alpha\beta \neq 0$, we obtain non-trivial dynamics for $\varphi$. 

\section{Black Holes}\label{BH}
We would like to investigate the implications of this model for black holes. Our approach is similar to  \cite{astorga2019compact}, in which metrics for black holes and other compact objects were studied for the general case of unimodular gravity with a non-conserved stress tensor, although here we focus on the black holes solutions of dynamical Henneaux-Teitelboim gravity in particular. For this, we consider the static, spherically symmetric ansatz
\bea  
ds^2 &=& - A(r) dt^2+ B(r) dr^2 + r^2 d\Omega^2_2, \label{symMetric}\\
 \mathbf{H} &=& h(r) \sin \theta \, dt \wedge d\theta \wedge d\phi, \label{symThreform}\\
 \varphi &=& \varphi(r), \label{symScalar}
\eea  
where the ansatz for $\mathbf{H}$ is the most general spherically symmetric, static three-form up to gauge transformations, and $d\Omega_2^2 = d\theta^2 + \sin^2\theta d\phi^2~$ is the round metric on the unit two-sphere. Additionally, we will only consider contributions to the stress tensor from $\mathbf{H}$ and $\varphi$, i.e., we will set $\mathcal{L}_M = 0$ in \eqref{dynamical action}. 

Since the stress tensor for $\mathbf{H}$ is pure trace, it will be convenient to divide the Einstein equations into trace-free and pure trace parts. The trace-free equations are \eqref{TFE}, while the pure-trace equation is
\be  
\varphi = \frac{1}{4}( R+\kappa T), \label{pure trace}
\ee  
where
\bea
R = \frac{1}{2 r^2 A^2 B^2}&\Big[&r^2 B A'^2+ r^2A A' B'-2rA B \left(r A''
+ 2 A'\right) \nonumber \\
&& \, + \, 4 A^2 \left(r B'+B^2-B\right)\Big]\label{R-sphere}
\eea
and
\be
T= \left(\beta \frac{\varphi'^2 }{B} - \frac{4 \alpha h'^2 }{AB r^4}\right) \label{T-sphere}
\ee
for the static, spherically symmetric ansatz \eqref{symMetric}-\eqref{symScalar}. \\ Similarly, the trace-free equations reduce to 
\begin{gather}
 u + \frac{1}{2rAB^2}(AB'+A'B) = \frac{-\kappa \beta}{4 B} (\varphi')^2 \label{TFE-tt} \\
  -u + \frac{1}{2rAB^2}(AB'+A'B) = -3\frac{\kappa \beta}{4 B} (\varphi')^2 \label{TFE-rr}\\
   u = \frac{\beta \kappa}{4B} (\varphi')^2 \label{TFE-angle-angle}
\end{gather}
where we have defined
\bea  
u = \frac{1}{8 r^2 A^2 B^2}\!\!\!&\Big(& \!\!2ABA''r^2-AB'A'r^2 -B(A')^2 r^2 \nonumber \\
&& \, + \, 4A^2B^2-4A^2B\Big). \label{uDef}
\eea  
In the above, \eqref{TFE-tt}, \eqref{TFE-rr} and \eqref{TFE-angle-angle} are the $tt$, $rr$ and $\theta\theta$ components of the trace-free Einstein equations, while the $\phi\phi$ component is identical to the $\theta\theta$ component, and all the off-diagonal components vanish. Equations \eqref{TFE-tt}-\eqref{TFE-angle-angle} are linearly dependent: by either subtracting \eqref{TFE-angle-angle} from \eqref{TFE-tt} or adding \eqref{TFE-angle-angle} to \eqref{TFE-rr}, we obtain
\be  
\frac{1}{2rAB^2}(AB'+A'B) = -\frac{\kappa \beta}{2 B} (\varphi')^2 \label{TFE-combo}
\ee    
which can be integrated to give
\be
\label{eqAB}
  AB = e^{-\kappa \beta \int_{r_0}^r \tilde{r} \varphi'(\tilde{r})^2 d \tilde{r}} \equiv e^{f(r)}.
\ee

Inserting the above into \eqref{TFE-angle-angle} and using \eqref{uDef} leads to
\begin{eqnarray}
\hspace{-.5cm} \frac{1}{4} A''+\frac{1}{8} \beta \kappa r A' \varphi '^2 -\frac{A }{2 r^2} + \frac{1}{2 r^2}\, e^{f(r)} =\frac{\beta \kappa}{4} (\varphi')^2 A. \label{theAbove}
\end{eqnarray}
Doing the same for the trace part of the Einstein equation \eqref{pure trace}, \eqref{R-sphere}, \eqref{T-sphere} and inserting the result into \eqref{theAbove}  gives
\begin{align}
\label{aeq1}
   -\frac{A'}{r}- A\left( \frac{1}{r^2}-\frac{\kappa \beta \varphi'^2}{2}\right) +\left( \frac{1}{r^2}-\varphi\right)e^{f(r)} = \frac{ \alpha \kappa}{r^4} h'^2,
\end{align}
which is a first-order linear equation for $A(r)$. The general solution is
\be
A(r)=  \frac{1}{r}e^{\frac{f(r)}{2}} \left(\int _{\ell}^re^{\frac{1}{2} f(R)} \left(1-G(R) R^2\right)dR-2m \right)
\label{metricfinal}
\ee
where the mass $m$ appears as an integration constant, the lower endpoint of integration $\ell = \sqrt{\frac{3}{\Lambda_0}}$, and $\Lambda_0$ is fixed by matching to the (A)dS-Schwarzschild metric in the ``decoupling limit," on which we expand below, and we have defined
\begin{gather}
G(r)= \varphi +\frac{ \alpha \kappa}{r^4} e^{-f(r)} h'^2.
\end{gather}
From \eqref{delta phi} and \eqref{delta H}, we notice that we can recover the (A)dS-Schwarzschild metric in the limit as $\beta \to 0$, $1/\alpha \to 0$ and  $1/(\alpha\beta) \to 0$, which we call the \textit{decoupling limit}. Notice that this limit does not reduce the order of the equation of motions for the scalar $\varphi$ \eqref{ca5} or for the three form $\mathbf{H}$ \eqref{ca6}. In this case, equation \eqref{delta phi} leads to $d\mathbf{H} = \mathbf{V\!ol}(g)$ and consequently $ d*d\mathbf{H}=0$ in equation \eqref{delta H}. Then equation \eqref{ca6} is redundant, while equation \eqref{ca5} becomes $d\square\varphi = 0$ for which $\varphi =$ constant is a consistent solution. When $d\mathbf{H} = \mathbf{V\!ol}(g)$, the stress energy tensor for $\mathbf{H}$, is just a constant times the metric, which can be absorbed into the cosmological constant. At the level of the action, we can integrate out $\mathbf{H}$ by inserting the solution to the equation of motion for $\varphi$ which is $d\mathbf{H} = \mathbf{V\!ol}(g)$. Then, the gauge kinetic term just becomes $S_\mathbf{H}=\alpha\int \mathbf{V\!ol}(g)$, which contributes as a cosmological constant term. 
We now turn our attention to the equations of motion \eqref{ca5}, \eqref{ca6} for $\mathbf{H}$ and $\varphi$. In order to make progress, through the decoupling limit, we replace
 $g_{\mu\nu}$ in equations \eqref{ca5}, \eqref{ca6} with the (A)dS-Schwarzschild metric, $g^{(0)}_{\mu\nu}$.
Now the equation of motion for $\mathbf{H}$ \eqref{ca6} can be integrated once to give 
\begin{gather}
\label{finalh}
    h(r) -4\alpha \beta A_0(r)\left( h''(r)-\frac{2h'(r)} {r}\right)=\frac{1}{3}r^3+C,
\end{gather}
where $A_0(r) = 1 -\frac{2m}{r} - \frac{\Lambda}{3}r^2$ is the metric function for an (A)dS-Schwarzschild black hole.
 This is a linear, inhomogeneous equation for $h(r)$, so the general solution is of the form 
 \be
 h(r) = h_0(r) + h_p(r),
 \ee 
 where $h_0(r)$ is the solution of the homogeneous equation and $h_p(r)$ is a particular solution. One can check that $h_p(r) = \frac{1}{3}r^3+C$ is a satisfactory particular solution.
The corresponding homogeneous equation can be solved using the method of Frobenius---i.e., we make a generalized power series ansatz
\be 
h_0(r) = \sum\limits_{n=0}^{\infty} c_n r^{-n+q}. \label{Frobeniush}
\ee
The allowed values of $q$ and coefficients $c_n$ are calculated in \ref{sec:Frobenius}. Since the equation is second order, we find two possible values for $q$ corresponding to the two linearly independent solutions of the homogeneous equation corresponding to \eqref{finalh}:
\be  
q = q_{\pm}\equiv \frac{3}{2}\pm \frac{3}{2} \sqrt{1-\frac{1}{3\Lambda \alpha \beta}}. \label{firstIndicial}
\ee  
Now the solution to the homogeneous equation becomes
\be  
h_0(r) = \sum\limits_{n=0}^{\infty} c^+_n r^{-n+q_+} + \sum\limits_{n=0}^{\infty} c^-_n r^{-n+q_-}, \label{Frobeniush2}
\ee  
where the coefficients are determined by the following base cases and recursion relation 
\bea
    c^\pm_1 &=& 0 \\
    c^\pm_2 &=& \frac{c^\pm_0 \ 4 \alpha \beta \ q_\pm(q_\pm-3)}{\frac{4\Lambda \alpha \beta}{3} (q_\pm-2)(q_\pm-5)+1} \\
    c^\pm_n &=& \frac{4\alpha \beta }{1+\frac{4\alpha\beta\Lambda}{3}(-n+q_\pm)(-n+q_\pm-3)}\times \nonumber \\
    &&\Big[c^\pm_{n-2} (-n+q_\pm+2)(-n+q_\pm-1) \\ 
    && -2m \ c^\pm_{n-3} (-n+q_\pm+3)(-n+q_\pm)\Big], \quad  n > 2. \nonumber
\eea
The scalar field equation of motion is given by
\begin{equation}\label{phieq}
    \varphi-4\alpha\beta\left(A'+\frac{2}{r}A\right)\varphi'-4\alpha\beta A\varphi'' =\Lambda,
\end{equation}
where $A$ and $\varphi$ both depend on the radius $r$. Again, we work around the Schwarzschild-(A)dS metric $A(r) = A_0(r)$, and apply the method of Frobenius, with our ansatz given by
\begin{equation}
    \varphi -\Lambda=\sum_{n=0}^{\infty}a_{n}r^{-n+s}.
\end{equation}
We once again refer the reader to appendix \ref{sec:Frobenius} for the computation of the allowed values of $s$ and  the coefficients $a_{n}$. The roots of the indicial equation are
\begin{equation}\label{sref}
    s=s_{\pm}\equiv -\frac{3}{2}\pm \frac{3}{2}\sqrt{1-\frac{1}{3\Lambda\alpha\beta}},
\end{equation}
so again, we have two independent solutions, as we would expect for a second order equation. Now the solution becomes
\be  
\varphi = \Lambda + \sum\limits_{n=0}^{\infty} a^+_n r^{-n+s_+} + \sum\limits_{n=0}^{\infty} a^-_n r^{-n+s_-}, \label{Frobeniush3}
\ee  
where the coefficients are
\begin{eqnarray}
    a^\pm_{1} &=& 0 \\
    a^\pm_{2} &=& \frac{a^\pm_{0}\left(2s_\pm +\frac{3}{4\Lambda\alpha\beta} \right)}{\frac{2\Lambda}{3}(2s_\pm+1)} \\
    a_{n}^{\pm} &=& \frac{1}{\frac{\Lambda}{3} (-n+s_\pm)(-n+s_\pm+3)+\frac{1}{4\alpha \beta}}\times \nonumber\\&&\Big[a^\pm_{n-2}(-n+s_\pm+2)(-n+s_\pm+3) \\&&-2m\ a^\pm_{n-3} (-n+s_\pm+3)^2\Big] , \quad n>2. \nonumber
\end{eqnarray}
Note that $\Re(s_\pm) < 0$ for all values of $\Lambda$, $\alpha$ and $\beta$, so that $\lim_{r\to \infty}\varphi(r) = \Lambda$, with $s_{\pm}$ being real for $\Lambda < 0$ or $\Lambda \geq \frac{1}{3 \alpha \beta}$. On the interval $0 < \Lambda < \frac{1}{3\alpha\beta}$, $s_\pm$ become complex, including the physically realistic case when $\Lambda$ is small and positive. One might worry that such a situation precludes a physically meaningful solution for the most relevant region of parameter space, however this is not the case. 
So far, we have been assuming that the integration constants $a^+_0$ and $a^-_0$ are real and independent, but they still define a solution of the homogeneous equation even when they are complex, albeit a generically complex-valued one. If we instead take $a^-_0=\Bar{a}^+_0$, then $\varphi$
is real when $s_\pm$ are complex conjugates. 
After some algebra, one finds the explicitly real form 
\bea
    \hspace{-.5cm}\varphi = \Lambda + \sum_{n = 0 }^\infty 2r^{-n+\Re(s_+)}&\!\!\!\big[&\!\!\! w_n \cos\big(\Im(s_+)\,\,\textrm{ln}(r)\,\big) \nonumber \\ 
    &&\,- \, z_n \sin\big(\Im(s_+)\,\textrm{ln}(r)\big)\big] .
\eea
where $w_n = \Re(a^+_n)$ and $z_n = \Im(a^+_n)$.
The series expansions above completely characterize the solutions for $\varphi$ and $\mathbf{H}$ in the asymptotic region. However, the series will not converge for all values of $r$. We will provide expressions for the radii of convergence of all of our series solutions at the end of this section. With this in mind, we now search for a solution to the equations of motion for small $r$. \\
The equation of motion for $h(r)$ is still given by \eqref{finalh}, but now we guess a power series of the form $h_{0}(r)=\sum\limits_{n=0}^{\infty}\tilde{c}_{n}r^{n+\tilde{q}}$. The indicial equation is simply $\tilde{q}(\tilde{q}-3)=0$, giving the two roots $\tilde{q}_1=3$ and $\tilde{q}_2=0$. In general, the full solution for the case of roots $\tilde{q}_1 > \tilde{q_2}$ differing by an integer is given by 
\be 
h_0(r) = \sum\limits_{n=0}^{\infty}\tilde{c}_{n}r^{n+\tilde{q}_1} + k \ln(r) \ \sum\limits_{n=0}^{\infty}\tilde{c}_{n}r^{n+\tilde{q}_1} +\sum\limits_{n=0}^{\infty}\tilde{d}_{n}r^{n+\tilde{q}_2},
\ee  
where the procedure for determining the coefficients $\tilde{d}_n$ depends on whether or not $k$ vanishes. If $k$ does vanish, the $\tilde{d}_n$ can be obtained using the usual recurrence relation, and otherwise they must be computed using a more complicated approach. In general, the value of $k$ is determined by 
\be  
k = \lim_{\tilde{q} \to \tilde{q}_2}(\tilde{q}-\tilde{q}_2)\tilde{c}_{\tilde{q}_1 - \tilde{q}_2}(\tilde{q}).
\ee  
For this particular differential equation, the recurrence relation for general $\tilde{q}$ is given by 
\begin{eqnarray}
    \Tilde{c}_{1}(\tilde{q}) &=& \tilde{c}_{2}(\tilde{q})=0 \\
    \Tilde{c}_{n+3}(\tilde{q}) &=& \frac{1}{{8 \alpha\beta m (n+3 +\tilde{q}) (n+\tilde{q})}}\times \nonumber \\ &&\Big[4\alpha\beta \tilde{c}_{n+2}(n+2+\tilde{q})(n-1+\tilde{q})\\
    &&-\tilde{c}_{n}(1+\frac{4\alpha\beta \Lambda}{3}(n+\tilde{q})(n-3+\tilde{q}))\Big], \quad n\geq 0, \nonumber
\end{eqnarray}
from which we conclude that 
\be  
k = \lim_{\tilde{q} \to 0}\tilde{q}\,\tilde{c}_{3}(\tilde{q}) = 0.
\ee 
Hence, the general solution becomes
\be 
h_0(r) = \sum\limits_{n=0}^{\infty}\tilde{c}_{n}r^{n+3} +\sum\limits_{n=0}^{\infty}\tilde{d}_{n}r^{n},
\ee  
where the coefficients are given by:
\begin{eqnarray}
   \tilde{c}_{1} &=& \tilde{c}_{2}=0 \\
    \Tilde{c}_{n+3} &=& \frac{4\alpha\beta \tilde{c}_{n+2}(n+5)(n+2)-\tilde{c}_{n}(1+\frac{4\alpha\beta \Lambda}{3}(n+3)n)}{8 \alpha\beta m (n+6)(n+3)},\nonumber \\
\end{eqnarray}
and
\begin{eqnarray}
    \Tilde{d}_{1} &=& \tilde{d}_{2}=0 \\
    \Tilde{d}_{n+3} &=& \frac{4\alpha\beta \tilde{d}_{n+2}(n+2)(n-1)-\tilde{d}_{n}(1+\frac{4\alpha\beta \Lambda}{3}n(n-3))}{8 \alpha\beta m (n+3) n},\nonumber \\
\end{eqnarray}
with $n\geq 0$.
The equation of motion for $\varphi(r)$ is given by \eqref{phieq}, but now we guess a power series of the form $\varphi(r) = \Lambda +\sum\limits_{n=0}^{\infty}\tilde{a}_{n}r^{n+\tilde{s}}$. The  indicial equation is
\begin{gather}
    \tilde{s}^2=0,
\end{gather}
giving repeated root $\tilde{s}_1=\tilde{s}_2=0$, while the recurrence relation for general $\tilde{s}$ is given by
 \begin{eqnarray}
    \Tilde{a}_{1}(\tilde{s}) &=&\tilde{a}_{2}(\tilde{s})= 0 \\
    \Tilde{a}_{n+3}(\tilde{s}) &=& \frac{1}{2m (n+\tilde{s}+3)^2}\times \nonumber\\ &&\Big[\tilde{a}_{n+2}(n+\tilde{s}+2)(n+\tilde{s}+3)\\&&-\tilde{a}_{n}\big(\tfrac{1}{4\alpha\beta}+\tfrac{ \Lambda}{3}(n+\tilde{s})(n+\tilde{s}+3)\big)\Big], \quad n\geq 0.\nonumber
\end{eqnarray} 
In the special case of a repeated root, the solution takes the general form 
\bea
\!\!\!\!\!\varphi(r) = \Lambda &+& \sum\limits_{n=0}^{\infty}\tilde{a}_{n}r^{n+\tilde{s}} \nonumber \\
&+& \tilde{K} \Big(\ln(r) \sum\limits_{n=0}^{\infty}\tilde{a}_{n}r^{n+\tilde{s}} +\sum\limits_{n=1}^{\infty}\tilde{b}_{n}r^{n+\tilde{s}}\Big)
\eea 
where $\tilde{K}$ is an integration constant. Here the coefficients $a_n$ are given by
\begin{eqnarray}
\Tilde{a}_{1} &=&\tilde{a}_{2}= 0 \\
    \hspace{-3cm} \Tilde{a}_{n+3} &=& \frac{\tilde{a}_{n+2}(n+2)(n+3)-\tilde{a}_{n}(\frac{1}{4\alpha\beta}+\frac{ \Lambda}{3}n(n+3))}{2m (n+3)^2},\nonumber\\
    \end{eqnarray}
and the coefficients $b_n$ are related to the $a_n$ by
    \begin{eqnarray}
    \tilde{b}_{n} &=&\frac{d}{d\tilde{s}}\tilde{a}_{n}(\tilde{s})\Big|_{\tilde{s}=0},
\end{eqnarray}
with $n\geq 0$.

The radius of convergence of each power series solution is the distance in the complex plane between the point around which the series is expanded and the nearest other regular singular point of the differential equation. For both \eqref{finalh} and \eqref{phieq}, the regular singular points are located at $r = 0$, $r = \infty$, and the zeros of $A_0(r)$. Thus, the radii of convergence are determined by the roots of a general depressed cubic equation 
\begin{gather}
    r^3-\frac{3}{\Lambda}r +\frac{6m}{\Lambda}=0
\end{gather}
with discriminant
\begin{gather}
    \Delta = \frac{1}{\Lambda^3}-\frac{9 m^2}{\Lambda^2}.
\end{gather}
The roots can be written as
\bea
    R_1 &=& S_+ +S_-\\
    R_\pm &=& -\frac{1}{2}R_1\pm\frac{1}{2}i\sqrt{3} (S_+-S_-)
\eea 
where
$$S_\pm=\sqrt[3]{-\frac{3m}{\Lambda}\pm\left(-\frac{1}{\Lambda^3}+\frac{9m^2}{\Lambda^2}\right)^{1/2}}.$$
If $\Lambda<0$, the discriminant is always negative and $R_1$ will be real, while $R_\pm$ are complex conjugates. The moduli are $|R_1|^2=(S_++S_-)^2 $ while $|R_\pm|^2={(S_+ +S_-)^2-3S_+S_-}$. Then, $S_+>0$ and $S_-<0$, leading to $|R_\pm|>|R_1|$. So the radius of convergence of the $r$-series is given by $0 < r < |R_1|$ while the radius of convergence of the $\frac{1}{r}$-series is given by $0<\frac{1}{r}<\frac{1}{|R_\pm|}$.

If $\Lambda>0 $, we have two cases. If $\Lambda>\frac{1}{9m^2}$  then the discriminant is negative and we still have $R_1$ real with $R_\pm$ being complex conjugates. Since $S_+ \!<0$ and $S_-\!<0$, $|R_1|>|R_\pm|$. Then the radius of convergence of the $r$-series is given by $0 < r < |R_\pm|$ while the radius of convergence of the $\frac{1}{r}$-series is given by $0<\frac{1}{r}<\frac{1}{|R_1|}$. 

Finally, if $0<\Lambda<\frac{1}{9m^2}$, then $S_+$ and $S_-$ become complex conjugates and we have three real roots. 
This can be seen by writing $S_\pm=(a\pm ib)^{1/3}= \rho^{1/3} e^{\pm i\theta/3}$, where $a=-\frac{3m}{\Lambda}$, $b=\left(\frac{1}{\Lambda^3}-\frac{9m^2}{\Lambda^2}\right)^{1/2}$, $\rho=\sqrt{a^2+b^2}$ and $\theta=\arctan^{-1}(b/a)$.
Then, $|R_1|=2 \rho^{1/3} \cos (\theta/3)$,  and $|R_\pm|=\rho^{1/3} (\cos (\theta/3) \pm \sqrt{3} \sin(\theta/3))$. Since $b/a < 0$, i.e. $-\pi/2 \leq \theta \leq 0$,  we have $|R_1| \geq |R_-| \geq |R_+|$ with equalities occurring only at the endpoints of the interval.
Hence, the radius of convergence of the $r$-series is given by $0 < r < |R_-|$ ,
while the radius of convergence of the $\frac{1}{r}$-series is given by  $0<\frac{1}{r}<\frac{1}{|R_1|}$.

With the series solutions for $\varphi$ and $\mathbf{H}$ in hand, one can now, take into account the approximate back-reaction on the (A)dS-Schwarzschild background by substituting the solutions for the fields into the general solution for the metric function \eqref{metricfinal}.
\begin{figure}[h]
    \centering
    \includegraphics[width=8.75cm]{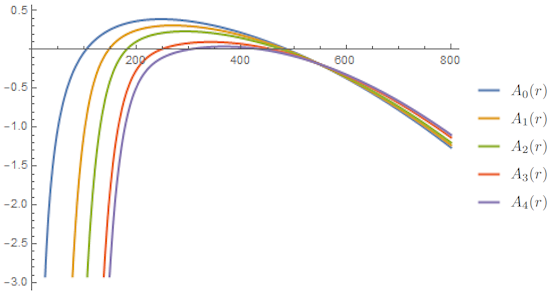}
    \caption{Plots of the metric function $A(r)$ are shown for various choices of the initial values $(\Bar{a}^+_0,\Bar{a}^-_0) =\{0,0.002,0.004,0.008,0.01\}\times(m_{\tiny{\astrosun}}^{-(s_+ + 2)},m_{\tiny{\astrosun}}^{-(s_- + 2) })$, where the label $A_i(r)$ coincides with the position of the value in the list, $m_{\tiny{\astrosun}}$ is the mass of the sun, and $s_+ \approx 0$, $s_- \approx -3$. Here we have fixed $m = 50\, m_{\tiny{\astrosun}}$,  $\Lambda_0=10^{-5}m_{\tiny{\astrosun}}^{-2}$, $r_0 = 10^5m_{\tiny{\astrosun}}$, $\beta = 10^{-3} m_{\tiny{\astrosun}}^4$, $\alpha = \frac{1}{8}\times 10^6 m_{\tiny{\astrosun}}^{-2}$ and chosen the particular solution $h(r) = \frac{1}{3}r^3.$ As $a^\pm_0 \to 0$, the solution approaches the dS-Schwarzschild metric $A_0(r)$, while for larger values of $a^\pm_0$ the event horizon is pushed outward while the cosmological horizon is pushed inward. At some critical values of $a^\pm_0$ the horizons degenerate and the black hole becomes extremal, while for still larger values of $a^\pm_0$ the solution has a naked singularity.}
    \label{fig:Aplot}
    \includegraphics[width=8.75cm]{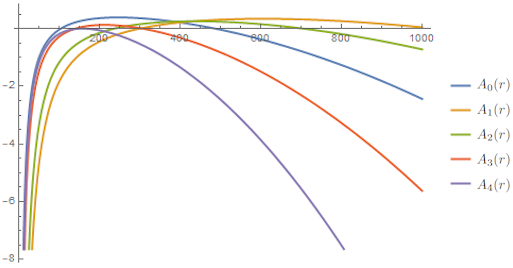}
    \caption{Plots of the metric function $A(r)$ are shown for various values of $\Lambda_0 =\{ 1.00, 0.25, 0.50,  2.00, 4.00\}\times 10^{-5}m_{\tiny{\astrosun}}^{-2}$, where the label $A_i(r)$ coincides with the position of the value in the list$, m_{\tiny{\astrosun}}$ is the mass of the sun, and we have fixed $a_0^{+}= .005m_{\tiny{\astrosun}}^{-(s_+ + 2)}, a_0^{-}= .005m_{\tiny{\astrosun}}^{-(s_- + 2)}$, with $s_+ \approx 0$, $s_- \approx - 3$, $m = 50 m_{\tiny{\astrosun}}$, $r_0 = 10^5m_{\tiny{\astrosun}}$, $\beta = 10^{-3}m_{\tiny{\astrosun}}^4$, $\alpha = \frac{1}{8}\times 10^6 m_{\tiny{\astrosun}}^{-2}$ and chosen the particular solution $h(r) = \frac{1}{3}r^3.$ We see that for small values of $\Lambda_0$ both horizons occur for larger values of $r$, while for larger values of $\Lambda_0$ the horizons are pushed inward and closer together until at some critical value, the horizons degenerate and the black hole becomes extremal. For even larger values of $\Lambda_0$, no horizons exist and the solution has a naked singularity.}
    \label{fig:Bplot}
\end{figure}

Using the series solutions for large $r$, one can perform a numerical integration to plot the profile function $A(r)$ whose zeros are the black hole and cosmological horizons for different values of $a^-_0$ and $\Bar{a}^+_0$, as shown in figure \ref{fig:Aplot}. As expected, for appropriate values of the couplings and integration constants, we find a pair of horizons which tend to the dS-Schwarzschild ones as $a^{\pm}_0 \to 0$. Note that the discrepancy from $A_0(r)$ of the inner horizon is greater than the outer horizon. However, as $\varphi$ is allowed to fluctuate more strongly, i.e., as  $a^\pm_0$ becomes large,  the solution does not have a horizon. In figure \ref{fig:Bplot} we show the results for different values of $\Lambda_0$. In order for the solution to approach dS-Schwarzschild in the decoupling limit, we fix $\Lambda_0 = \Lambda + \alpha\kappa$. For $\Lambda_0 < 0$, the lower endpoint of integration in \eqref{metricfinal} becomes imaginary and the solution becomes complex and unphysical. Hence, we see that  the model has a  dynamical preference for black holes with $\Lambda_0 \geq 0$.

This completes our analysis of spherically symmetric black holes in dynamical Henneaux-Teitelboim gravity.
Another very natural question to consider is how the dynamics of the effective cosmological constant modify the standard cosmological background of general relativity, which we study in the following section. 
\section{Cosmological analysis}\label{FRW}
We will now consider homogeneous, isotropic configurations sourced only by the fields $\varphi$ and $\mathbf{H}$ with no additional matter content. Our analysis runs along the same lines as other work on cosmology in unimodular gravity with a non-conserved stress tensor, e.g. \cite{garcia2019cosmic}, where the model we consider here serves as a concrete realization of the general proposal. 

We consider the FRW metric
\begin{equation}\label{metric 2}
    ds^{2}=-dt^{2}+a(t)^{2}(dx^{2}+dy^{2}+dz^{2})
\end{equation}
along with the homogeneous and isotropic fields
\begin{eqnarray}\label{H 2}
    \mathbf{H} &=& h(t)\,dx\wedge dy\wedge dz, \\
    \varphi &=& \varphi(t).
\end{eqnarray}
Again, we are only considering contributions to the stress tensor from $\mathbf{H}$ and $\varphi$, so we set $\mathcal{L}_{M}=0$ in \eqref{dynamical action}. 

We decompose the Einstein equations into trace-free and pure trace parts, recalling that the stress tensor for $\mathbf{H}$ is pure trace. The trace-free equation is given by \eqref{TFE}, and plugging in our symmetry-reduced expressions for the metric, three-form, and scalar, we find a single independent equation:
\begin{equation}
    \frac{\kappa}{2}\beta\dot{\varphi}^{2}=-\frac{\dot{a}^{2}-a\ddot{a}}{a^{2}}. \label{Friedmann1}
\end{equation}
The trace equation \eqref{pure trace} gives
\be
    \varphi=\frac{3}{2}\left(\frac{\ddot{a}}{a}+\frac{\dot{a}^{2}}{a^{2}} \right)-\frac{\kappa}{4}\left(\beta\Dot{\varphi}^{2}+\alpha\frac{\Dot{h}^{2}}{6a^{6}} \right),\label{Friedmann2}
\ee
which can be combined with \eqref{Friedmann1} to give
\begin{equation}
    \varphi=\frac{\Ddot{a}}{a}+2\frac{\dot{a}^{2}}{a^{2}}-\frac{\kappa\alpha}{24}\frac{\dot{h}^{2}}{a^{6}}.\label{FriedmannFinal}
\end{equation}
The equations of motion for the scalar and three-form imply \eqref{ca5} and \eqref{ca6}, which upon symmetry reduction become
\begin{gather}\label{H EoM}
    \dot{h} -4\alpha\beta\frac{d}{dt}\left(a^{3}\frac{d}{dt}\Big(\frac{\dot{h}}{a^{3}}\Big)\right)=a^{3} \\
\label{Phi EoM}
    \varphi-4\alpha\beta\left(3\frac{\dot{a}}{a}\dot{\varphi}+\ddot{\varphi}\right)=\Lambda.
\end{gather}
Making use of the Hubble parameter $H=\frac{\dot{a}}{a}$, defining $u=\frac{\varphi}{\Lambda}$, $v=\frac{\dot{h}}{a^{3}}$, and adimensionalizing, we simplify our problem to three equations\footnote{It is worth noting the close similarity between equations \eqref{sr1}, \eqref{sr2} and \eqref{sr3} and the slow roll equations. See e.g. \cite{liddle1994formalizing}.}:
\begin{gather}
    u-1=3\tilde{H}\dot{u}+\ddot{u} \label{sr1}\\
    v -1=3\dot{v}\tilde{H}+\ddot{v} \label{sr2}\\
    \tilde{\Lambda} u=\dot{\tilde{H}}+3\tilde{H}^{2}+pv^{2}, \label{sr3}
\end{gather}
where $\tilde\Lambda = 4\alpha\beta\Lambda$, $p=-\kappa\alpha^{2}\beta/24$, $\tilde{H} = \sqrt{4\alpha\beta}H$, and $\tilde{t} = t/\sqrt{4\alpha\beta}$. Thus, we search for a fixed point, and find one at $u=1$, $v=1$, and $\tilde{H}=H_{0}\equiv\sqrt{\frac{\tilde{\Lambda}-p}{3}}$. Perturbing around this fixed point decouples two of the equations, yielding
\begin{gather}\label{}
    \delta u=3H_{0}\delta\dot{u}+\delta\ddot{u} \\
\label{}
    \delta v=3H_{0}\delta\dot{v}+\delta\ddot{v} \\
    \tilde{\Lambda}\delta u-2p\delta v=\delta\dot{\tilde{H}}+6H_{0}\delta \tilde{H}.
\end{gather}
These equations are solvable. The first two equations simply give
\begin{gather}
    \delta u=A_{u}e^{\omega_{+}\tilde{t}}+B_{u}e^{\omega_{-}\tilde{t}} \\
    \delta v=A_{v}e^{\omega_{+}\tilde{t}}+B_{v}e^{\omega_{-}\tilde{t}}
\end{gather}
where $\omega_{\pm}=-\frac{3H_{0}}{2}\pm\sqrt{\frac{9}{4}H_{0}^{2}+1}$. The last equation has a homogeneous solution given by
\begin{equation}
    \delta \tilde{H}^{(h)}=C_{H}e^{-6H_{0}\tilde{t}}
\end{equation}
and a particular solution given by
\begin{equation}
    \delta \tilde{H}^{(p)}=A_{H}e^{\omega_{+}\tilde{t}}+B_{H}e^{\omega_{-}\tilde{t}},
\end{equation}
where the coefficients are
\begin{gather}
    A_{H}=\frac{\tilde\Lambda A_{u}-2p A_{v}}{\omega_{+}+6H_{0}} \\
    B_{H}=\frac{\tilde\Lambda B_{u}-2p B_{v}}{\omega_{-}+6H_{0}}.
\end{gather}
If we impose initial conditions such that $A_u = A_v = 0$, which imply that $A_H = 0$, then $\varphi$ relaxes to a constant at late times, allowing for a de Sitter-like phase of accelerated expansion in the late universe. However, in general both the exponentially growing and decaying modes will be present, so we expect that the solution that is compatible with observations of our current cosmological epoch will be unstable to small perturbations around this background. Still, having a solution of the linearized equations of motion which approaches de Sitter at late times may be useful for more sophisticated analyses that include matter other than $\mathbf{H}$ and $\varphi$. 

\section{Discussion}\label{Discus}
In this work we have introduced dynamical Henneaux-Teitelboim gravity, a minimal modification of general relativity based closely on the Henneaux-Teitelboim formulation of unimodular gravity that realizes fluctuations in the cosmological constant sourced by non-conservation of the energy-momentum tensor. Furthermore, we have considered the implications of this model for spherically symmetric black holes and homogeneous, isotropic configurations. In order to make progress analytically, we have made a number of simplifying assumptions that should ultimately be lifted in order to probe more realistic physical scenarios. For the analysis of black holes, we have only presented an approximate solution of all the field equations by perturbing around the Schwarzschild-de Sitter metric and using series methods. In order to establish whether such an approximate solution is physically viable we would need, at a minimum, to study the geodesic motion of a massive particle orbiting such a black hole around the equator and compare with astrophysical observations. The assumption of spherical symmetry dramatically simplifies the equations of motion, reducing them to ordinary differential equations, but of course, in order to model rotating black holes, this restriction must be lifted. Studies of rotating black holes in similar modified gravity models such as dynamical Chern-Simons gravity \cite{alexander2009chern, yagi2012slowly, yunes2009dynamical, alexander2021chern, alexander2022black1} suggest that going to the slowly-rotating limit and making use of numerical methods are likely to be necessary. Finally, the series solutions we have presented for small and large $r$ have finite radii of convergence, leaving the possibility of intermediate regions, possibly containing the event horizon, which are not covered by either series solution. Here again, numerical methods for interpolating between the solutions for small and large $r$ might be useful. We leave such numerical studies for the future, but note that for the solutions we have studied in the decoupling limit that are close to Schwarzschild-de Sitter, the radius of convergence is well below the scale of the event horizon.
On the cosmology side, we have only considered the contribution to the stress tensor from the fields $\mathbf{H}$ and $\varphi$ for simplicity. In order to make realistic predictions, we would need, at a minimum, to include contributions from general baryonic matter and radiation as well, consider the effects of cosmological perturbations, and, if they are stable, study the matter-power spectrum for deviations from $\Lambda$CDM. We have also seen that while a solution of the linearized equations of motion that approaches de Sitter at late times exists, this solution is apparently unstable. It would be interesting to see whether this instability persists in a more realistic analysis including additional matter and radiation, or whether, perhaps, the exponentially growing modes might be dynamically suppressed by the inclusion of additional matter degrees of freedom.
The model we have considered in this article is a natural extension of the Henneaux-Teitelboim formulation of unimodular gravity because the kinetic terms for the scalar and three-form fields are invariant under the local symmetries of the latter---spacetime diffeomorphisms and two-form-valued gauge transformations. Consequently, the kinetic terms are forced upon us by ``Gell-Mann's totalitarian principle," which dictates that ``everything that is not forbidden (by local symmetries) is compulsory." To put it plainly, renormalization group flow will generate every term that is compatible with the local symmetries of the action \eqref{HTaction}, and that includes kinetic terms for the scalar and three-form fields. In this sense, the model we have studied in this article follows straightforwardly from \eqref{HTaction} and general principles of effective field theory.

\section{Acknowledgements}
We are grateful to Tucker Manton and Roman Marcarelli for discussions and helpful feedback on an early version of this manuscript. EA was supported by the STFC Consolidated Grant ST/W507519/1.

\appendix 
\section{Coefficients and Indices for Series Solutions}\label{sec:Frobenius}
In this section we perform detailed calculations of the coefficients and indices for the series solutions of the equations of motion for $\mathbf{H}$ and $\varphi$ in the spherically symmetric case. Let us first focus on the homogeneous equation
\be 
\label{h-homogeneous}
    h_0(r) -4\alpha \beta A_0(r)\left( h_0''(r)-\frac{2h_0'(r)} {r}\right) = 0.
\ee  
Putting the series ansatz $h_0(r) = \sum\limits_{n=0}^{\infty} c_n r^{-n+q}$ into \eqref{h-homogeneous} gives
\begin{align}
     \sum\limits^{\infty}_{n=0} \Big[c_n  r^{-n+q}-4\alpha \beta c_n (-n+q)(-n+q-3)\nonumber \\ \left(r^{-n+q-2}-2m \ r^{-n+q-3}-\tfrac{\Lambda}{3}r^{-n+q}\right)\Big]=0.
\end{align}
Now, reindexing
\bea  
   &\sum\limits^{\infty}_{n=0}& \!\! c_n (1+\tfrac{4\alpha\beta\Lambda}{3}(-n+q)(-n+q-3))  r^{-n+q} \nonumber \\
   - 4\alpha \beta \!\!\!&\sum\limits^{\infty}_{n=2}&\!\! c_{n-2} (-n+q+2)(-n+q-1) r^{-n+q}\\
   - 4\alpha \beta \!\!\! &\sum\limits^{\infty}_{n=3}& \!\!(-2m)c_{n-3} (-n+q+3)(-n+q) r^{-n+q}=0 \nonumber
\eea  
Now, let us find the recurrence relation and the allowed values of $q$.
Firstly, consider the base cases.
For $n=0$ we have only terms of order $r^q$, which yields the indicial equation
\begin{gather}
   c_0 \left(1+\frac{4\Lambda \alpha \beta}{3} q(q-3)\right)=0.
\end{gather}
Hence, 
\begin{gather}
    c_0=0 \quad \textrm{or}\quad q = q_{\pm}\equiv \frac{3}{2}\pm \frac{3}{2} \sqrt{1-\frac{1}{3\Lambda \alpha \beta}}.
\end{gather}
We can discard $c_0 = 0$ without loss of generality, since this amounts to reindexing the sum to start from $n = 1$.
For $n=1$, we have only terms of order $r^{q-1}$, so
\begin{gather}
   c_1 \left(1+\frac{4\Lambda \alpha \beta}{3} (q-1)(q-4)\right)=0,
\end{gather}
which tells us that either
\begin{gather}
    c_1=0 \quad \textrm{or}\quad q = \frac{5}{2}\pm \frac{3}{2} \sqrt{1-\frac{1}{3 \Lambda \alpha \beta}}.
\end{gather}
We accept the first solution since the second one is inconsistent with $c_0 \neq 0$.
For $n=2$, we have only terms of order $r^{q-2}$, so
\begin{gather}
    c_2 \left(1+\frac{4\alpha \beta \Lambda}{3} (q-5)(q-2)+\right)-c_0 4\alpha \beta q(q-3)=0,
\end{gather}
which tells us that
\begin{eqnarray}
    c_2 &=& \frac{c_0 \ 4 \alpha \beta \ q(q-3)}{\frac{4\Lambda \alpha \beta}{3} (q-2)(q-5)+1}
\end{eqnarray}
Then, for $n\geq 3$ we have only terms of order $r^{q-n}$, so
\begin{gather}
    c_n=\frac{4\alpha \beta }{1+\frac{4\alpha\beta\Lambda}{3}(-n+q)(-n+q-3)}\times \nonumber \\ \Big[c_{n-2} (-n+q+2)(-n+q-1) \\-2m \ c_{n-3} (-n+q+3)(-n+q)\Big]\nonumber
\end{gather}
We now work with the scalar field equation
\begin{equation}
    \varphi-4\alpha\beta\left(A'\varphi'+\frac{2}{r}A\varphi'+A\varphi'' \right)=\Lambda
\end{equation}
Subtracting $\Lambda$ from both sides and letting $y=\varphi-\Lambda$:
\begin{equation}
    A(r)y''(r)+\left(\frac{2}{r}A(r)+A'(r) \right)y'(r)-\frac{1}{4\alpha\beta}y(r)=0
\end{equation}
Noting the Schwarzschild-AdS metric $A(r)=1-\frac{2m}{r}-\frac{\Lambda}{3}r^{2}$, we have:
\begin{equation}
\label{finalphi}
    \left(1 - \frac{2m}{r} - \frac{\Lambda}{3}r^2\right) y''+\left(\frac{2}{r}-\frac{2m}{r^2}-\frac{2\Lambda}{3}r\right) y'-\frac{1}{4 \alpha \beta} y=0
\end{equation}
Plugging in the series ansatz $y=\sum\limits_{n=0}^{\infty}a_{n}r^{-n+s}$ yields:
\begin{eqnarray}
   &\sum\limits^{\infty}_{n=0}& a_n (-n+s)(-n+s+1)  r^{-n+s-2} \nonumber \\
   -\!\!\!&\sum\limits^{\infty}_{n=0}&\ 2m\  a_n (-n+s)^2 r^{-n+s-3} \\
   -\!\!\!&\sum\limits^{\infty}_{n=0}&a_n\left( (-n+s)(-n+s+3)\frac{\Lambda}{3}+\frac{1}{4\alpha\beta}\right) r^{-n+s}=0 \nonumber
\end{eqnarray}
Reindexing gives:
\begin{eqnarray}
   &\sum\limits^{\infty}_{n=2}& a_{n-2} (-n+s+2)(-n+s+3)  r^{-n+s} \nonumber \\ 
   -\!\!\!&\sum\limits^{\infty}_{n=3}&\ 2m \ a_{n-3} (-n+s+3)^2 r^{-n+s} \\
   -\!\!\!&\sum\limits^{\infty}_{n=0}&a_n \left((-n+s)(-n+s+3)\frac{\Lambda}{3}+\frac{1}{4\alpha\beta}\right) r^{-n+s}=0  \nonumber
\end{eqnarray}
Now we focus on the recurrence relation. First consider the base case $n=0$. Here we only have terms of order $r^{s}$, so we have:
\begin{equation}
    a_0 \left(\frac{\Lambda}{3} s(s+3)+\frac{1}{4\alpha \beta}\right)=0.
\end{equation}
Hence
\begin{equation}
    a_0=0 \quad \textrm{or}\quad s = s_{\pm}\equiv -\frac{3}{2}\pm \frac{3}{2} \sqrt{1-\frac{1}{3\Lambda \alpha \beta}}.
\end{equation}
We can discard $a_{0}=0$ without loss of generality since this amounts to reindexing the sum to start from $n=1$. For $n=1,$ we have only terms of order $r^{s-1}$, so we have
\begin{equation}
   a_1 \left(\frac{\Lambda}{3} (s-1)(s+2)+\frac{1}{4\alpha \beta}\right)=0,
\end{equation}
Thus we have 
\begin{equation}
    a_1=0 \quad \textrm{or}\quad s = s_{\pm} + 1.
\end{equation}
The second solution is inconsistent with $a_{0}\neq 0$. For $n=2,$ we have only terms of order $r^{s-2}$, giving us:
\begin{equation}
   - a_2 \left(\frac{\Lambda}{3} (s+1)(s-2)+\frac{1}{4\alpha \beta}\right)+a_0(s+1)s=0,
\end{equation}\
and so we find
\begin{eqnarray}
    a_2 &=& \frac{a_0(s+1)s}{\frac{\Lambda}{3} (s+1)(s-2)+\frac{1}{4\alpha \beta}} \nonumber \\
    &=& \frac{a_0\left(2s+\frac{3}{4\Lambda\alpha \beta }\right)}{\frac{2\Lambda}{3} (2s+1)},
\end{eqnarray}
where we used $s^{2}+3s=-\frac{3}{4\Lambda\alpha\beta}$. Finally, for $n\geq 3$, we have terms of order $r^{s-n}$, so we have:

\begin{eqnarray}
    &&-a_n \left(\tfrac{\Lambda}{3} (-n+s)(-n+s+3) \nonumber+\tfrac{1}{4\alpha \beta}\right)\nonumber \\
    &&+ a_{n-2}(-n+s+2)(-n+s+3) \\
    &&- 2m \,a_{n-3}(-n+s+3)^3=0 \nonumber
\end{eqnarray}
which gives us the recurrence relation

\begin{eqnarray}
    a_{n} &=& \frac{1}{\frac{\Lambda}{3} (-n+s)(-n+s+3)+\frac{1}{4\alpha \beta}}\times \nonumber\\&&\Big[a^\pm_{n-2}(-n+s+2)(-n+s+3) \\&&-2m\ a_{n-3} (-n+s+3)^2\Big] , \quad n>2. \nonumber
\end{eqnarray}

\bibliographystyle{ieeetr}
\bibliography{bib}

\end{document}